\documentclass[aps,prx,amsmath,amssymb,twocolumn,showpacs,superscriptaddress, longbibliography]{revtex4-2}
\usepackage[pdftex]{graphicx}
\usepackage[colorlinks=true,citecolor=blue]{hyperref}
\usepackage{braket}
\usepackage{graphicx,textcomp,amssymb,amsmath,dcolumn,hyperref,upgreek}
\usepackage[utf8]{inputenc}

\usepackage{blindtext}

\begin{document}

\title{Post-processing of real-time quantum event measurements for an optimal bandwidth}

\author{J.~Kerski}
\email{jens.kerski@uni-due.de}
\author{H.~Mannel}
\author{P.~Lochner}
\author{E.~Kleinherbers}
\affiliation{Faculty of Physics and CENIDE, University of Duisburg-Essen, Lotharstr. 1, 47057 Duisburg, Germany}
\author{A.~Kurzmann}
\affiliation{2nd Institute of Physics, RWTH Aachen University, 52074 Aachen, Germany}
\author{A.~Ludwig}
\author{A.~D.~Wieck}
\affiliation{Chair for Applied Solid State Physics, Ruhr-Universit{\"a}t Bochum, Universit{\"a}tsstr. 150, 44780 Bochum, Germany}
\author{J.~König}
\author{A.~Lorke}
\author{M.~Geller}
\affiliation{Faculty of Physics and CENIDE, University of Duisburg-Essen, Lotharstr. 1, 47057 Duisburg, Germany}
               
\date{\today}

\begin{abstract}
	Single electron tunneling and its transport statistics have been studied for some time using high precision charge detectors. However, this type of detection requires advanced lithography, optimized material systems and low temperatures (mK). A promising alternative, recently demonstrated, is to exploit an optical transition that is turned on or off when a tunnel event occurs. High bandwidths should be achievable with this approach, although this has not been adequately investigated so far. We have studied low temperature resonance fluorescence from a self-assembled quantum dot embedded in a diode structure. We detect single photons from the dot in real time and evaluate the recorded data only after the experiment, using post-processing to obtain the random telegraph signal of the electron transport. This is a significant difference from commonly used charge detectors and allows us to determine the optimal time resolution for analyzing our data. We show how this post-processing affects both the determination of tunneling rates using waiting-time distributions and statistical analysis using full-counting statistics. We also demonstrate, as an example, that we can analyze our data with bandwidths as high as 350\,kHz. Using a simple model, we discuss the limiting factors for achieving the optimal bandwidth and propose how a time resolution of more than 1\,MHz could be achieved. 
\end{abstract}

\maketitle

% - - - - - - - - - - - - - - - - - - - - - - - - - - - - - - - - - - - - - -

\section{Introduction}

The fast sensing of single quantum events is an important topic for many areas of modern physical research. For example, in the fabrication of ever smaller transistor structures, the charge quantization is becoming increasingly important \cite{DiVincenzo1995}. Also, a deeper understanding of quantum transport \cite{Blanter2000} could give new insights for current issues, such as the fast readout of various qubit candidates \cite{Shnirman1998, Jang2021} or quantum sensing \cite{Degen2017, Kerski2021}. The tunneling of single electrons can be studied using the random telegraph signal (RTS), where the state of a quantum system is measured in real-time and the quantum mechanical fluctuations can be statistically evaluated \cite{Bagrets2003, Kleinherbers2018, Sifft2021}.

The corresponding experiments are mainly performed by electrostatic sensing of gate-defined quantum dots using quantum point contacts (QPCs) and single-electron transistors (SETs). Bandwidths of 10-100\,kHz (for QPCs) \cite{Gustavsson2006, Wagner2019} and up to 1\,MHz (for SETs) \cite{Lu2003} have already been achieved. However, these electrical readout techniques of the quantum state are limited to low temperatures (mK), require advanced lithography, and are difficult to transfer to other physical systems. In contrast to this electrical detection, the RTS of an optical transition, which is switched "on" or "off" by the tunneling electrons, can also be used to read out the quantum state of the system. This optical detection is not tied to a specific material system or to a specific optical excitation/detection method. For example, time-resolved photoluminescence can be used to study charge fluctuations in colloidal nanocrystals at room temperature \cite{Galland2011, Zbydniewska2015}. Also, resonant fluorescence at compared to the mK scale high temperatures (4.2\,K) can be used to study tunneling \cite{Kurzmann2019} and Auger processes \cite{Lochner2020} in self-assembled quantum dots (QDs) as well as spin dynamics \cite{Vamivakas2010} in quantum dot molecules. The optical excitation of the single quantum emitter by resonant fluorescence offers a high energetic resolution, and the detection of the single photons with avalanche photodiodes or superconducting nanowire detectors promises a high bandwidth. In the case of single photon detection, the time stamp of each photon is recorded during the experiment and only afterwards the RTS is determined by post-processing of these time stamps. To achieve a similar resolution in electrical measurements, would require to measure the detector current with single electron resolution, which has not been possible so far.

In this work, we demonstrate a post-processing procedure on single photon streams to obtain the optimal bandwidth in the RTS of electron tunnel events between a reservoir and a self-assembled quantum dot. We show how the time resolution affects both the determination of the tunneling rates involved and the evaluation by full-counting statistics. We demonstrate examplarily that we can evaluate our data with up to 350\,kHz bandwidth. We discuss the limitation of our method using a simple model, compare it to our data, and propose how a time resolution of more than 1\,MHz can be achieved.

\section{Method}
\label{sec:method}

\begin{figure}[tb]
	\includegraphics[width=\linewidth]{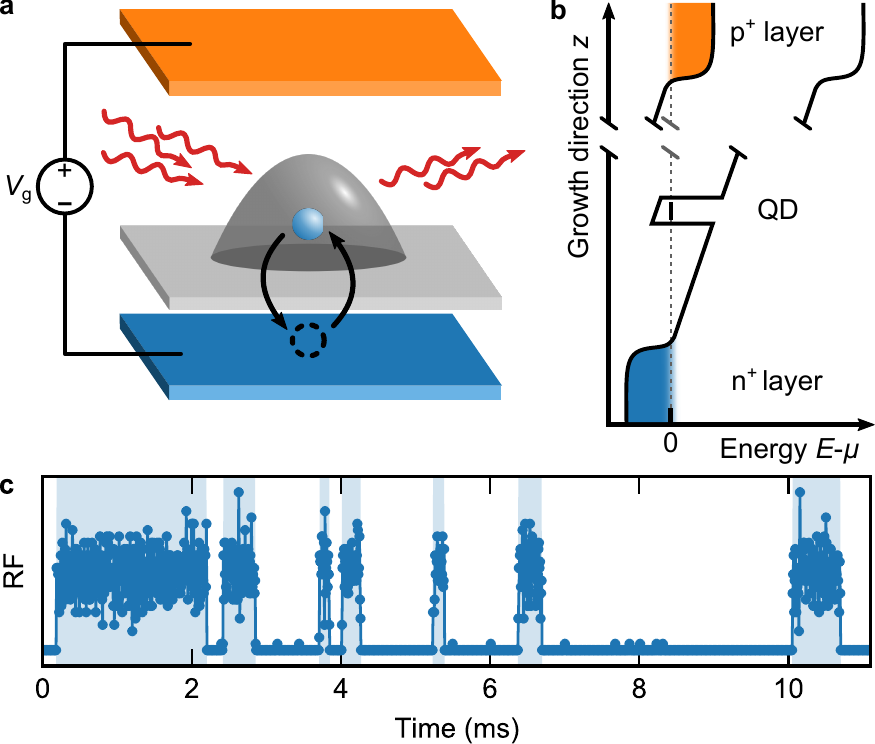}
	\caption{\textbf{Sample structure and random telegraph signal.} {\bf a}, The QD is resonantly-excited by a tuneable laser and embedded in a pin-diode structure. Its charge state can be tuned by applying a voltage $V_\mathrm{g}$ to the diode that sets the ground state of the QD in resonance with the chemical potential of the $\mathrm{n}^+$-doped back contact. {\bf b}, Schematic illustration of the conduction and valence band edge in respect to the chemical potential $\mu$ (dashed line). {\bf c}, The resonance fluorescence signal of the QD is switching between "on" and "off" due to electrons tunneling between the QD and the charge reservoir.}
	\label{fig1}
\end{figure}

We have studied single photon streams from a single self-assembled InAs/GaAs quantum dot.  The dot is embedded in a p-i-n diode structure (see Fig.~\ref{fig1}(a), (b) and \cite{Lochner2020} for more details). The $\mathrm{n}^+$-doped back contact serves as an electron reservoir, separated from the QD by a tunnel barrier of 30\,nm GaAs, 10\,nm $\mathrm{Al}_{0.33}\mathrm{Ga}_{0.67}\mathrm{As}$ and 5\,nm GaAs. By applying a voltage $V_\mathrm{g}$ to the diode, it is possible to bring the QD states into resonance with the chemical potential in the reservoir, so that an electron can tunnel between the dot and the reservoir. The sample is mounted in a bath cryostat at $T=4.2$\,K. By carefully adjusting $V_\mathrm{g}$, we can position the quantum dot level within the stepwidth of the Fermi-Dirac distribution in the reservoir ($2k_BT\approx700$\,µeV). As a result, both the time-averaged occupancy of the quantum dot and the tunneling in/out rates involved can be precisely controlled.

Resonance fluorescence (RF) is used to detect the dot's charge state. A laser is tuned into resonance with the neutral exciton transition of the empty dot, while it is out of resonance for the singly charged dot. To suppress the scattered laser photons we use the cross-polarisation technique \cite{Vamivakas2009, Matthiesen2012, Kurzmann2016a}. The single photons are detected by an avalanche photodiode (average jitter: 300 ps) and recorded as time stamps by a time-to-digital converter with a time resolution of 81 ps (see \cite{Kurzmann2019} for more details). The switching between the two charge states results in a switching between high and low detected photon rates. This is the random telegraph signal (see Fig.~\ref{fig1}(c)).

\begin{figure}[t!]
	\includegraphics[width=\linewidth]{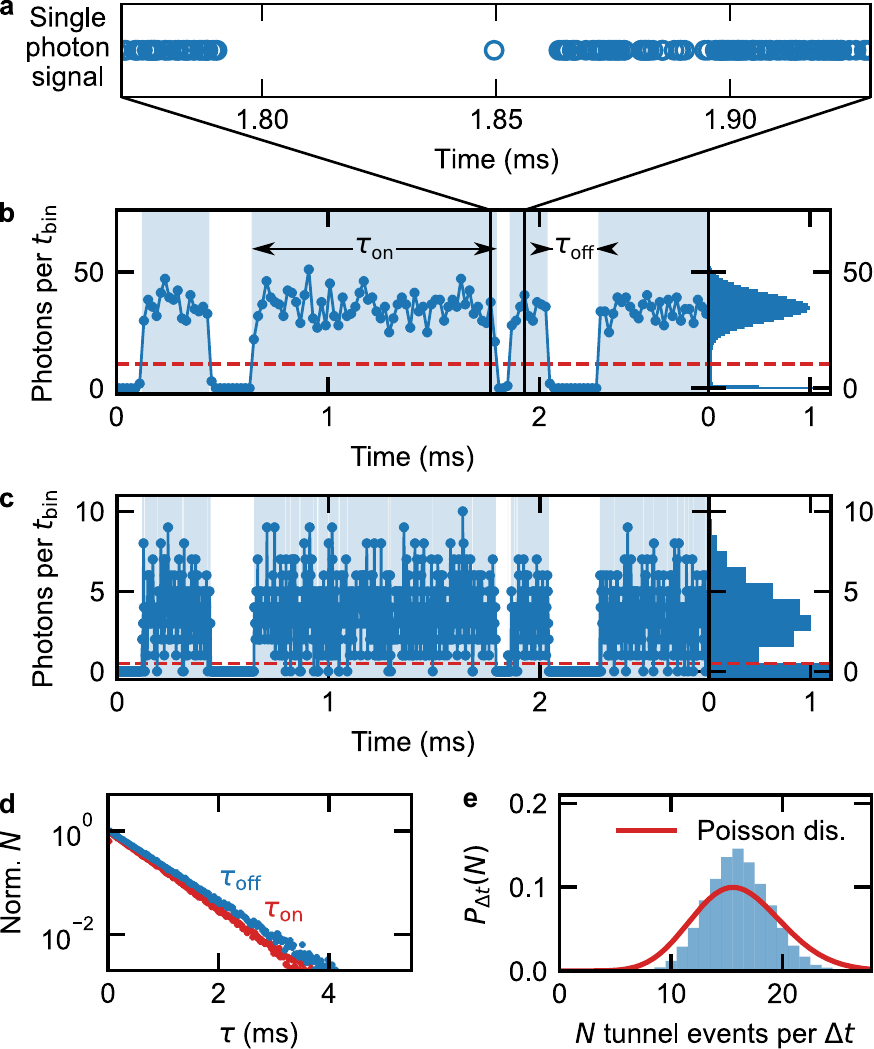}
	\caption{\textbf{Evaluation scheme of the random telegraph signal (RTS).} {\bf a}, Single photon stream from the resonant excitation of the exciton transition, recorded for about $5$\,min {\bf b, c}, Binned photon counts per $t_\mathrm{bin}$ (left panel) and corresponding histogram (right panel) for two binning times $t_\mathrm{bin}$ ({\bf b}: $t_\mathrm{bin}=20$\,µs; {\bf c}: $t_\mathrm{bin}=2$\,µs). The charge state of the dot is determined by applying a threshold photon number (dashed red line). {\bf d}, The waiting-time distribution is evaluated by the length and frequency of the time periods of the on ($\tau_\mathrm{on}$) and off state ($\tau_\mathrm{off}$), respectively. {\bf e}, The full-counting statistics of the data in Fig.~\ref{fig2}(b), i.e. the probability to find $N$ tunnel events in a given time period $\Delta t=20$\,ms.}
	\label{fig2}
\end{figure}

 The photon stream (see Fig.~\ref{fig2}(a)) was typically recorded for 5\,min and holds the information about the charge state of the QD, but consists of discrete time stamps for every detected photon. To evaluate the photon stream and obtain the random telegraph signal (RTS) of the tunnel events, we need a time trace for the number of photons per time interval. We achieve this by choosing a binning time $t_\mathrm{bin}$ and counting the time stamps of the detected photons in this interval for the entire photon stream. As the time interval $t_\mathrm{bin}$ is the smallest time unit in the time trace of the RTS, the inverse binning time $1/t_\mathrm{bin}$ corresponds to the highest frequency that can be detected. In the following, we will also refer to the quantity $1/t_\mathrm{bin}$ as the binning rate, which can also be understood as the highest achievable bandwidth. However, it should be pointed out here that this bandwidth is chosen \emph{after} the measurement and is not predetermined during data acquisition. 

Fig.~\ref{fig2}(b) shows an example of such a binned signal for $t_\mathrm{bin}=20$\,µs, where two levels ("on" and "off") of the photon count rate can be clearly distinguished. As mentioned above, these levels correspond to the two charge states of the quantum dot: The exciton transition can be excited (high photon rate, on state) for an empty dot, while the transition is quenched (low photon rate, off state) for a charged dot. The levels are also visible in a histogram of the signal (right panel). The photon counts of the on and off state are Poisson distributed. The few photon counts observed for the off state are dark counts from the avalanche photodiode and spurious laser photons reflected from the sample, which have not interacted with the QD. The charge state of the dot is now determined from this binned time trace in Fig.~\ref{fig2}(b) by applying a threshold photon number (dashed red line). Every data point above this threshold is assigned to the on state and every data point below it to the off state. This results in a binary signal (blue shaded background color in Fig.~\ref{fig2}(b) and (c)).

As mentioned above, we choose the binning rate after the experiment. However, it is not useful to choose an arbitrarily high binning rate, as the number of detected photons is limited and the distributions of the on and off state overlap in that case (see Fig.~\ref{fig2}(c) with $t_\mathrm{bin} = 2$\,µs). Therefore, we define the optimal bandwidth as the highest binning rate at which the dynamics of the electron transport can still be properly determined from the data. How to find the optimal bandwidth in a post-processing procedure is the topic in the following part of this paper.

The binary signal (blue shaded background color in Fig.~\ref{fig2}(b) and (c)) can be evaluated in different ways. On the one hand, by evaluating the waiting-time distribution (WTD) \cite{Brandes2008}, i.e. the frequency with which a given time period $\tau_\mathrm{on}$ or $\tau_\mathrm{off}$ occurs. Here $\tau_\mathrm{on}$ and $\tau_\mathrm{off}$ are the intervals during which the QD is on or off. In our measurements, the WTDs can be described by single exponentials, with decay rates, that correspond to the tunneling in and out rates $\gamma_\mathrm{in}$ and $\gamma_\mathrm{out}$, respectively (see Fig.~\ref{fig2}(d)). Since we are only interested in the decay rates, we present the frequencies normalized to their extrapolated value for $\tau=0$.

On the other hand, the tunnel events themselves can be evaluated by full-counting statistics (FCS) \cite{Levitov1996, Kleinherbers2018, Bagrets2003}. For FCS, we determine the number of tunnel events in a given time interval $\Delta t$. We study the longtime limit and therefore the interval must be significantly larger than the inverse tunneling rates. Since our tunneling processes are in the sub-ms range, we choose $\Delta t = 20$\,ms. The result is a distribution that can be analyzed using statistical moments and cumulants \cite{Kurzmann2019, Gustavsson2006, Stegmann2015, Kleinherbers2021}. As an example, the blue curve in Fig.~\ref{fig2}(e) shows the probability to find $N$ tunnel events in a 20\,ms time interval, for the data presented in Fig.~\ref{fig2}(b). Comparison with a calculated Poissonian (red line) reveals that the distribution is sub-Poissonian.

\section{Results and Discussion}

\begin{figure}[t!]
	\includegraphics[width=\linewidth]{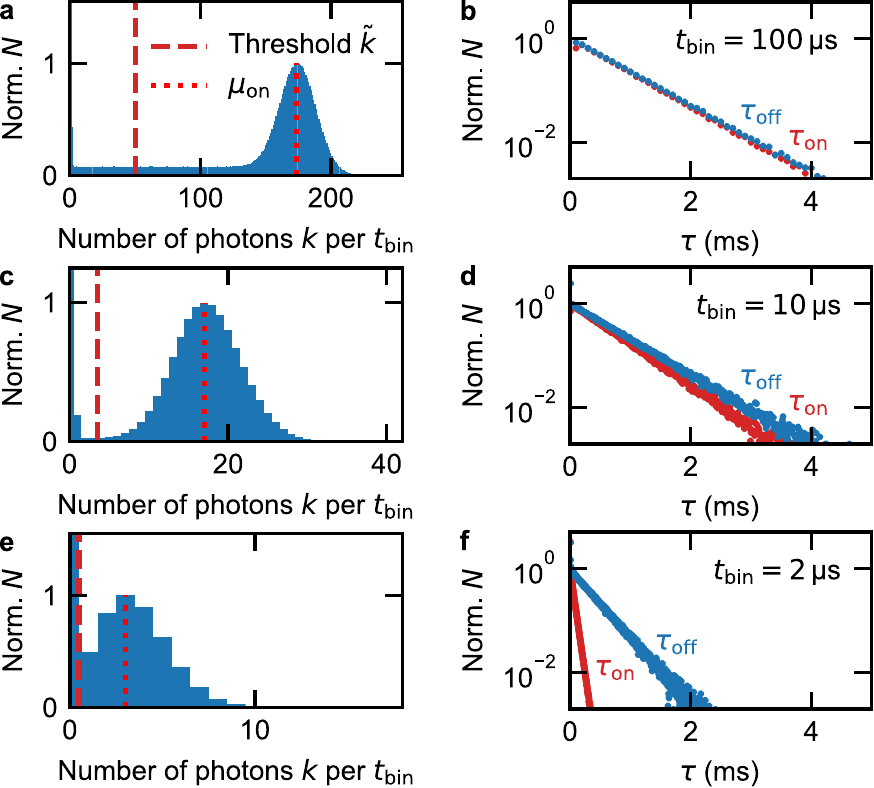}
	\caption{\textbf{Histograms and waiting-time distributions for different $t_\mathrm{bin}$.} {\bf a, c, e}, Histograms for $t_\mathrm{bin}=100$\,µs, 10\,µs and 2\,µs, respectively. The histograms exhibit the on and off state distributions with a constant background. The choice of the threshold photon number $\tilde{k}$ (red dashed) separates the on from the off state. {\bf b, d, f}, Corresponding waiting-time distributions. The time spans $\tau_\mathrm{on}$ and $\tau_\mathrm{off}$ show an exponential decay associated with the tunneling rate into and out of the quantum dot. For 100\,µs and 10\,µs the tunneling rates are very similar. For 2\,µs the rates differ strongly as increasingly more data points are assigned to the wrong QD state.}
	\label{fig3}
\end{figure}

In order to demonstrate how the optimal bandwidth can be obtained from the random telegraph signal, we have chosen for the binning time $t_\mathrm{bin}$ 13 logarithmically distributed values between 2\,µs and 100\,µs. Figs.~\ref{fig3}(a), (c) and (e) show the histograms of the binned photon counts $k$ for $t_\mathrm{bin}$ values of 100\,µs, 10\,µs and 2\,µs. They all exhibit Poisson distributions for both, the on and the off state. The corresponding expectation values are $\mu_\mathrm{on}\approx$ 175, 17, 3.6 (red dotted lines Figs.~\ref{fig3}(a), (c) and (e)) and $\mu_\mathrm{off}\ll1$ photons per $t_\mathrm{bin}$. We have normalized the histograms in each case with the value at $\mu_\mathrm{on}$. There is also a constant background in between. This can be seen for example for $t_\mathrm{bin} = 100$\,µs (see Fig.~\ref{fig3}(a)), where the background is $\approx0.1$ between about 1 and 120 photons per $t_\mathrm{bin}$. It reflects those time intervals in which a tunnel event takes place. Since the binning intervals are not correlated with the tunnel events, their numbers of photons are equally distributed between $\mu_\mathrm{on}$ and $\mu_\mathrm{off}$ \cite{Kleinherbers2021}.

The choice of the threshold photon number $\tilde{k}$ (red dashed line in Figs.~\ref{fig3}(a), (c) and (e)) is not critical for high values of $t_\mathrm{bin}$, because the distributions of the on and off states are well separated (see Fig.~\ref{fig3}(a)). Additionally, for $t_\mathrm{bin}=100$\,µs, the $\tau_\mathrm{on}$ and $\tau_\mathrm{off}$ values exhibit very low noise, see Fig.~\ref{fig3}(b). From both these facts we conclude that the evaluation of the data in Fig.~\ref{fig3}(a) and (b) is accurate and reliable. The fitted rates are $\gamma_\mathrm{in}\approx 1.53\,\mathrm{ms}^{-1}$ and $\gamma_\mathrm{out}\approx 1.48\,\mathrm{ms}^{-1}$. These are very similar, so both processes equally contribute to the electron transport between the dot and the reservoir. This is usually quantified by the asymmetry of the tunnel coupling $A = (\gamma_\mathrm{out}-\gamma_\mathrm{in})/(\gamma_\mathrm{out}+\gamma_\mathrm{in})$ \cite{Bagrets2003, Gustavsson2006}. For $t_\mathrm{bin}=100$\,µs, we find $A\approx-0.018$.

When we evaluate the same experimental data with decreasing $t_\mathrm{bin}$ in order to improve the time resolution, it becomes increasingly difficult to separate the distributions of the on and off states (see Fig.~\ref{fig3}(c) and (e)). Thus, the evaluation becomes less and less reliable. 

For $t_\mathrm{bin}=10$\,µs, $\tau_\mathrm{on}$ and $\tau_\mathrm{off}$ start to visibly differ (see Fig.~\ref{fig3}(d)), resulting in an asymmetry $A\approx-0.074$. Finally, for 2\,µs binning time, Fig.~\ref{fig3}(f), they differ so strongly, that the derived asymmetry $A\approx-0.72$ is close to the maximum possible value of $\left|A\right|=1$. This is due to the increasing overlap of the on and off state distributions (left column of Fig.~\ref{fig3}), as the overlap causes data points in the binned time trace to be on the wrong side of the threshold and, thus, be assigned to the wrong state. Such a false assignment will break up an otherwise continuous $\tau_\mathrm{on}$ interval into two shorter ones an thus skew the statistical distribution towards higher rates $\gamma_\mathrm{on}$. The same holds for the $\tau_\mathrm{off}$ statistics.

The evaluation of the WTD for 100\,µs in Fig.~\ref{fig3}(b) is not influenced by wrongly assigned data points and represents the correct tunneling rates in the system. Therefore we use the rates determined there and especially the asymmetry $A\approx0$ as a criterion for the optimal choice of the threshold photon number $\tilde{k}$ for shorter binning times. Based on this determination of $\tilde{k}$, we can now proceed to assess how well the transport statistics can be determined for a given $t_\mathrm{bin}$ using full counting statistics.

\begin{figure*}[t]
	\includegraphics[width=\linewidth]{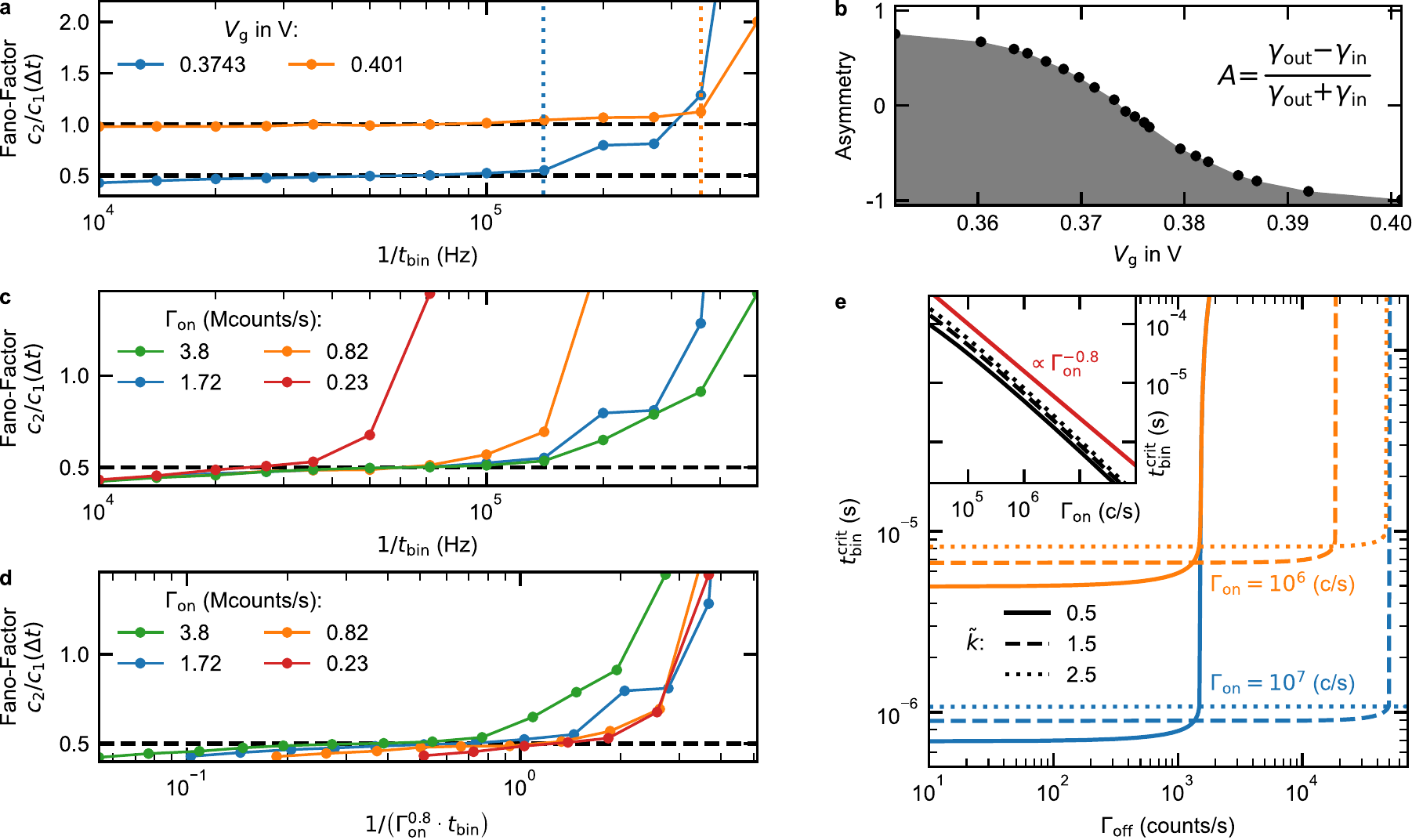}
	\caption{\textbf{Fano factor as a function of $1/t_\mathrm{bin}$.} {\bf a}, The Fano factor in dependence of $t_\mathrm{bin}$ at different bias voltages $V_\mathrm{g}$. The Fano factor is used here as very sensitive indicator for an appropriate choice of the binning rate (bandwidth), as the Fano factor corresponds to the expected value for low binning rates and diverges for higher rates. The transport statistics can be evaluated up to 140\,kHz (0.3742\,V; dotted blue line) and 350\,kHz (0.401\,V; dotted orange line). {\bf b} The asymmetry as a function of bias voltage is given by the tunneling rates that were determined in the waiting time distributions in Fig.~\ref{fig2}. {\bf c}, The strong increase of the Fano factor can be shifted by changing the laser intensity and thus the average photon rate of the on state $\Gamma_\mathrm{on}$. {\bf d}, By rescaling the binning rate, the results for 0.23\,Mcounts/s, 0.82\,Mcounts/s, and 1.72\,Mcounts/s coincide. The data for 3.8\,Mcounts/s are slightly shifted. {\bf e}, The critical binning time $t_\mathrm{bin}^\mathrm{crit}$, as obtained from the model, versus the mean photon rate of the off state. (Inset) The dependence of $\Gamma_\mathrm{on}$ can be described by $t_\mathrm{bin}^\mathrm{crit}\propto\Gamma_\mathrm{on}^{-0.8}$ (inset). The threshold photon number $\tilde{k}$ at which the maximum binning rate can be obtained depends on the average photon rate of the off state $\Gamma_\mathrm{off}$ (see solid, dashed and dotted lines).}
	\label{fig4}
\end{figure*}

We characterize our probability distributions $P_{\Delta t}(N)$ by the Fano factor $f=c_2/c_1$, where $c_i$ is the $i$-th cumulant of $P_{\Delta t}(N)$, more specifically $c_1$ is the mean and $c_2$ is the variance. The Fano factor reflects whether the transport is limited by a single tunnel process ($f=1$; Poisson-distributed $P_{\Delta t}(N)$) or by two tunnel processes ($f=0.5$; sub-Poissonian distributed $P_{\Delta t}(N)$) \cite{Gustavsson2006}. The more complex evaluation of the higher order cumulants, as well as the factorial cumulants is possible \cite{Kurzmann2019, Kleinherbers2021}, however, this evaluation is not relevant here and we will concentrate on the first and second cumulant and the Fano factor. 

For the measurement presented in Fig.~\ref{fig3}, both tunneling processes contribute equally to the transport. Accordingly, we expect a Fano factor of $0.5$. In Fig.~\ref{fig4}(a) (blue data points), we show the evaluated Fano factors for the already mentioned 13 logarithmically distributed binning rates. All blue data points were evaluated from the same single-photon stream recorded at a bias voltage of $V_\mathrm{g} = 0.3743$\,V. For low binning rates, the Fano factor is very close to the expected value. However, above about 140\,kHz (dotted blue line), the Fano factor diverges. To show that this divergence is reproducible and not only applies to the described asymmetry of $0$, we used the bias voltage $V_\mathrm{g}$ to change the tunneling rates \footnote{The transition energy of the neutral exciton shifts due to the quantum-confined Stark effect \cite{Fry2000}. To compensate for this, we applied a magnetic field of 2\,T in the growth direction to lock the transition to the incident laser by 'dragging', due to coupling to the nuclear spin bath of the quantum dot \cite{Latta2009, Xu2009, Hoegele2012, Urbaszek2013}.}. For $V_\mathrm{g}=0.401$\,V, the tunneling rate into the dot is significantly faster than before ($\gamma_\mathrm{in}\approx2.9\,\mathrm{ms}^{-1}$) but the tunneling rate out of it is strongly suppressed ($\gamma_\mathrm{out}\approx42.5\,\mathrm{s}^{-1}$) and, thus, an asymmetry $A\approx-1$ is obtained (see Fig.~\ref{fig4}(b)). In this situation, where the transport is limited by only one process, we expect a Fano factor which equals unity. The corresponding data (orange data points in Fig.~\ref{fig4}(a)) shows qualitatively the same diverging behavior as for a Fano factor of 0.5. It also shows that we can resolve the single-electron transport statistics with a binning rate of about 350\,kHz (dotted orange line).

The question to be clarified now is how to maximize the binning rate in order to increase the optimum bandwidth. Therefore, we will discuss in the following why the increase of the Fano factor occurs, when it occurs and which parameters influence it. As already discussed in connection to the WTD in Fig.~\ref{fig3}, high binning rates lead to an incorrect state assignment for some data points. This manifests itself not only in shortened time intervals $\tau_\mathrm{on}$ and $\tau_\mathrm{off}$ but also in falsely detected, additional tunnel events. These have no physical meaning, so we refer to them as binning-induced tunnel events in the following. When these binning-induced events outnumber the correct tunnel events, the Fano factor starts to diverge and an evaluation of the real transport statistics is no longer possible.

Two main approaches are now promising to increase the time resolution: (1) Somewhat counter-intuitively, this can be achieved in systems with faster dynamics (higher tunneling rates). More tunnel events per $\Delta t$ would allow a faster evaluation, since the same number of binning-induced events could be better compensated by more correct events. This approach could be realized on similar sample structures with thinner tunneling barriers. (2) Another approach is to reduce the overlap between the on and off state distributions. In our optical method, the off state is already close to the technical limit, where the observed photon events are exclusively given by the dark counts of the APD. The mean value of the on state distribution $\mu_\mathrm{on}$, however, can be shifted to higher values, e.g. by increasing the laser intensity or optimizing the detection efficiency. To demonstrate this, we have repeated the measurement for asymmetry $A=0$ with a total of four different laser intensities (see Fig.~\ref{fig4}(c)). We denote the different data sets according to the average rate of detected photons of the on state $\Gamma_\mathrm{on}$. The previously presented data set (blue data points) corresponds to $\Gamma_\mathrm{on}\approx1.72\,\mathrm{MCounts}/\mathrm{s}$.

At low binning rates ($\leq30$\,kHz) in Fig.~\ref{fig4}(c), the determined Fano factor for all laser intensities is almost identical and close to 0.5, indicating that the different excitation of the QD does not affect the system dynamics. This is remarkable since in electrical measurements with quantum point contacts \cite{Young2010} and radio-frequency single electron transistors \cite{Aassime2001} as detectors, a backaction on the investigated system is observed. Figure~\ref{fig4}(c) also shows that the critical binning rate $1/t_\mathrm{bin}^\mathrm{crit}$, at which the Fano factor starts to diverge, increases monotonically with the laser intensity. This can be seen, for example, by comparing the measurement with $\Gamma_\mathrm{on}\approx0.23\,\mathrm{MCounts}/\mathrm{s}$ (red data points in Fig.~\ref{fig4}(c)), which diverges already at about $35$\,kHz, with the data set for $\Gamma_\mathrm{on}\approx1.72\,\mathrm{MCounts}/\mathrm{s}$ (blue data points in Fig.~\ref{fig4}(c)), which does not diverge until $150$\,kHz. This is in agreement with our discussion about the overlap of the on and off state distributions.

In order to also quantitatively test our reasoning and enable statements about the possible scalability of the method, we consider a simple model for the divergence of the Fano factor for high $1/t_\mathrm{bin}$. It starts from the premise that an evaluation of the probability distribution $P_{\Delta t}(N)$ will definitely no longer be meaningful when the mean number of correct $N_\mathrm{corr}$ and binning-induced events $N_\mathrm{false}$ are equal. $N_\mathrm{corr}$ is given by the effective tunnel rate 
\begin{equation*}
	1/\gamma_\mathrm{tun}=1/\gamma_\mathrm{in}+1/\gamma_\mathrm{out}
\end{equation*}

 and the FCS time interval $\Delta t$. The number of binning-induced events $N_\mathrm{false}$ contains the misassigned on and off state data points
 \begin{equation*}
 	N_\mathrm{false}=N_\mathrm{false}^\mathrm{on}+N_\mathrm{false}^\mathrm{off}.
 \end{equation*}

The number $N_\mathrm{false}^\mathrm{on}$ is the product of the number of data points that should be assigned to the on state $N^\mathrm{on}$ and the probability that one of these data points is wrongly assigned $P^\mathrm{on}_\mathrm{false}$:
\begin{equation*}
N_\mathrm{false}^\mathrm{on}=P^\mathrm{on}_\mathrm{false}N^\mathrm{on}.
\end{equation*}
An analogous relation applies to the off state.

For an asymmetry $A=0$, the numbers $N^\mathrm{on}$ and $N^\mathrm{off}$ are equal ($N^\mathrm{on} = N^\mathrm{off} = \Delta t/(2t_\mathrm{bin}^\mathrm{crit})$). To obtain the probability $P^\mathrm{on}_\mathrm{false}$, we start from a Poisson distribution $P_{\mu_\mathrm{on}}(k)$ for the number of photons $k$ per $t_\mathrm{bin}$. The expectation value of the distribution is given by $\mu_\mathrm{on}=\Gamma_\mathrm{on}\cdot t_\mathrm{bin}^\mathrm{crit}$. Then, $P^\mathrm{on}_\mathrm{false}$ is given by the fraction of $P_{\mu_\mathrm{on}}(k)$ that falls below the threshold photon count $\tilde{k}$ and is therefore falsely assigned to the off state:
\begin{equation*}
	P_\mathrm{false}^\mathrm{on}=\sum_{k=0}^{\tilde{k}}P_{\mu_\mathrm{on}}(k).
\end{equation*}
An analog derivation holds for the off state. Combining the above equations, we find
\begin{equation}
	\gamma_\mathrm{tun}\Delta t =\sum_{k=0}^{\tilde{k}}P_{\mu_\mathrm{on}}(k)\frac{\Delta t}{2t_\mathrm{bin}^\mathrm{crit}} +\sum_{k=\tilde{k}}^{\infty}P_{\mu_\mathrm{off}}(k)\frac{\Delta t}{2t_\mathrm{bin}^\mathrm{crit}}.
\end{equation}

This relation does not have a simple analytical solution for $t_\mathrm{bin}^\mathrm{crit}$, therefore we have solved it numerically for an effective tunnel rate $\gamma_\mathrm{tun} \approx 0.75\,\mathrm{ms}^{-1}$ and various combinations of $\tilde{k}$, $\Gamma_\mathrm{on}$, and $\Gamma_\mathrm{off}$. For $\Gamma_\mathrm{off}<10^3$\,counts/s, both $\Gamma_\mathrm{off}$ and $\tilde{k}$ have little effect on the critical binning rate calculated from our model (see Fig.~\ref{fig4}(e)). In the experiments, a small value of $\Gamma_\mathrm{off}$ corresponds to a low detector dark count rate and well suppressed scattered laser light intensity. 

The dependence of $t_\mathrm{bin}^\mathrm{crit}$ on the average photon rate of the on state is found to be $t_\mathrm{bin}^\mathrm{crit}\propto \Gamma_\mathrm{on}^{-0.8}$ (see Fig.~\ref{fig4}(e) inset). Rescaling the laser-power-dependent data in Fig.~\ref{fig4}(c) using this power law, shows that the curves for 0.23\,Mcounts/s, 0.82\,Mcounts/s and 1.72\,Mcounts/s all coincide (see Fig.~\ref{fig4}(d)). Only the measurement with 3.8\,Mcounts/s (green line in Fig.~\ref{fig4}(d)) is an exception as here the increase of the Fano factor is shifted towards smaller rescaled binning rates. This shift can be attributed to the fact that the condition above, $\Gamma_\mathrm{off}<10^3$\,counts/s, is no longer fulfilled due to laser light scattered to the detector. Indeed, $\Gamma_\mathrm{off}$ for this data set is about 1.3\,kcounts/s, very close to the value, where an abrupt increase of $t_\mathrm{bin}^\mathrm{crit}$ is observed in Fig.~\ref{fig4}(e). Similarly, the data set for $\Gamma_\mathrm{on}=1.72$\,Mcounts/s (blue data points in Fig.~\ref{fig4}(d)) has a spurious off state count rate of 1.4\,kcounts/s, which may explain why one of the data points ($2\times10^5$\,Hz) does not fall onto the scaled curves. For the data sets with 0.23 and 0.82\,Mcounts/s, $\Gamma_\mathrm{off}$ is very low with 0.12 and 0.19\,kcounts/s, respectively, and thus in a range where $\Gamma_\mathrm{off}$ has no influence on the critical binning rate. Figure~\ref{fig4}(e) shows that the value of $\Gamma_\mathrm{off}$, where the critical binning time starts to diverge, is not fixed. For an experimental situation, where a high $\Gamma_\mathrm{off}$ is unavoidable, the divergence point can be shifted to much higher values of $\Gamma_\mathrm{off}$ by a proper choice of $\tilde{k}$, while only moderately increasing the critical binning time $t_\mathrm{bin}^\mathrm{crit}$.

Our optical measurements, their analysis and interpretation show that in contrast to an electrical measurement of a random telegraph signal, the single photon detection offers an optimum bandwidth that can be determined in post-processing, i.e. after the actual experiment in the lab.

We have shown that a critical binning rate and thereby optimal bandwidth exists, above which the evaluation of the random telegraph signal is not valid anymore. The simple model we have considered shows that the reason for this is the overlap of the photon distributions of the on and off states. This occurs when the binning rate is too large and leads to a misassignment of the states. In our experiment, this overlap is enhanced by technical imperfections, such as dark counts of the photodetector, reflected laser photons and an imperfect photon collection rate. However, even for an ideal measurement, the increasing overlap with decreasing binning time would impose a limit on the maximum time resolution and thus represents an absolute physical limit for this method.

\section{Conclusion and Outlook}
We have measured single photon streams of a resonantly excited optical transition from a self-assembled quantum dot to study the random telegraph signal of electron tunneling between the dot and a reservoir. We have shown that the data obtained by this approach has an inherent maximum time resolution and that the corresponding optimal bandwidth for analysis can be determined by a post-processing procedure. We have established a model that suggests a purely physical limit of this method due to the Poissonian nature of the detected photons. However, this limit can be increased using a stronger optical excitation in order to achieve higher photon rates. In the present experiment, we were able to evaluate single electron transport statistics with a bandwidth of 350 kHz. In other work, where resonator systems were used to increase the emission rate, photon rates of 40 MHz had been demonstrated \cite{Tomm2021}. Our results indicate that with such a system an evaluation of the random telegraph signal with a bandwidth of several MHz is possible.

\section*{Acknowledgment}

This work was funded by the Deutsche Forschungsgemeinschaft (DFG, German Research Foundation) – Project-ID 278162697 – SFB 1242 and the individual research grant No.~GE2141/5-1 and LU2051/1-1. Ar.~L. and A.~D.~W. acknowledge gratefully support of DFG-TRR-160, BMBF-Q.Link.X 16KIS0867, and the DFH/UFA CDFA-05-06.

% - - - - - - - - - - - - - - - - - - - - - - - - - - - - - - - - - - - - - -


\end{document}